\documentclass[prb,twocolumn,showpacs,superscriptaddress]{revtex4-1}

\usepackage{graphicx,amssymb,amsmath,color,bbm,psfrag}
\usepackage[utf8]{inputenc}
\usepackage{color}
\usepackage[normalem]{ulem}

\begin{document}
\definecolor{darkgreen}{rgb}{0,0.5,0}

\newcommand{\jav}[1]{\textcolor{red}{#1}}

\title{Anomalous hyperfine coupling and nuclear magnetic relaxation in Weyl semimetals}
\author{Zolt\'an Okv\'atovity}
\affiliation{Department of Theoretical Physics and BME-MTA Exotic  Quantum  Phases Research Group, Budapest University of Technology and Economics, Budapest, Hungary}

\author{Ferenc Simon}
\affiliation{Department of Physics and MTA-BME Lend\"{u}let Spintronics Research Group (PROSPIN),  Budapest University of Technology and Economics}

\author{Bal\' azs D\' ora}

\email{dora@eik.bme.hu}

\affiliation{Department of Theoretical Physics and BME-MTA Exotic  Quantum  Phases Research Group, Budapest University of Technology and Economics, Budapest, Hungary}

\date{\today}

\begin{abstract}
The electron-nuclear hyperfine interaction shows up in a variety of phenomena including e.g. NMR studies of correlated states and spin decoherence effects in quantum dots. 
Here we focus on the hyperfine coupling and the NMR spin relaxation time, $T_1$ in Weyl semimetals.
Since the density of states in Weyl semimetals varies with the square of the energy around the Weyl point, a naive power counting predicts
a $1/T_1T\sim E^4$ scaling with $E$ the maximum of temperature ($T$) and chemical potential.
By carefully investigating the hyperfine interaction between nuclear spins and Weyl fermions, we find that while its spin part behaves conventionally, its orbital part diverges unusually with the inverse of energy around the Weyl point. Consequently,
the nuclear spin relaxation rate scales in a graphene like manner as $1/T_1T\sim E^2\ln(E/\omega_0)$ with $\omega_0$ the nuclear Larmor frequency.
This allows us to identify an effective hyperfine coupling constant, which is tunable by gating or doping, which is relevant for decoherence effect in spintronics devices and double quantum dots
 where hyperfine coupling is the dominant source of spin-blockade lifting.
\end{abstract}

\pacs{76.60.-k,85.75.-d,03.65.Vf}

\maketitle

\section{Introduction}

Topological phenomena have percolated into condensed matter once again after the theoretical prediction\cite{bernevig} and experimental realization\cite{konig} of topological
insulators. Although their bulk is insulating similarly to a normal insulator, their surface hosts symmetry protected topological surface states, whose properties are determined
by topological invariants. This gives rise to the quantized spin-Hall conductivity in spin-Hall insulators\cite{hasankane,zhangrmp}
as well as topological spin textures, the topological magnetoelectric effect. In addition, the search for topological superconductors and Majorana fermions
has also received a significant boost.

The descendant of topological insulators in 3D 
is a Weyl semimetal\cite{herring,wan,murakami,BurkovPRL2011}, which could be also called a topological metal. This is characterized by
monopole like structures in momentum space, which come in pairs, and are protected by topology. Unlike their two dimensional counterparts, e.g. the Dirac cones
in graphene\cite{rmpguinea}, which appear at high symmetry points and can be easily gapped away by e.g. breaking the sublattice symmetry, these three
dimensional structures appear at non-symmetry protected points in the Brillouin zone and hence are robust against small perturbations and can only 
be annihilated when two monopoles with opposite topological charge collide into each other.

Weyl semimetals also feature a variety of peculiar phenomena, such as an anomalous Hall conductivity in 3D, whose "quantization" is proportional to the separation of the Weyl nodes in
momentum space\cite{BurkovPRL2011}. The chiral anomaly, i.e. the  anomalous non-conservation of an otherwise conserved quantity, the chiral current in this case,
has also been addressed experimentally\cite{chiral1,chiral2} after a wealth of theoretical papers.
Due to the non-trivial topology, the two monopoles in momentum space induce surface states, known as Fermi-arcs\cite{fermiarc1,fermiarc2}.
Weyl points also exist in artificially created band structures, e.g. in photonic crystals\cite{soljacic}.

In condensed matter physics, however, many other detection tools are at our disposal to probe materials at various energy scales. Among these, the nuclear magnetic resonance (NMR) has long been used\cite{winter,abragam,SlichterBook}
to unveil the nature of exotic states of matter.
In particular, NMR spectroscopy was found to be a useful diagnostic tool in revealing the nature and symmetry of pairing in superconductors\cite{HebelSlichter,maeno}.
At the heart of the NMR is the hyperfine coupling, i.e. the interaction between nuclear spin and surrounding conduction electrons.
In addition,  quantum information processing and spintronics relies
on long spin relaxation and coherence times of electrons in the devices. It is known that strong hyperfine effects can lead to decoherence thus limiting the device performance\cite{FabianRMP}.

\begin{figure}[t!]
\includegraphics[width=5cm]{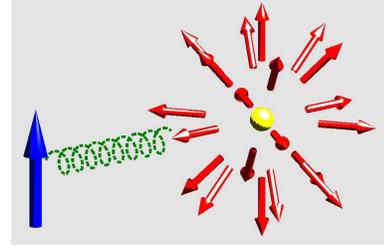}
\caption{(Color online) Cartoon of the hyperfine interaction (green spring) between a nuclear spin (blue arrow) and Weyl semimetal (red hedgehog structure with 
gold monopole inside). 
The radius of the hedgehog is set by the chemical potential.}
\label{nmrabra}
\end{figure}

In general, the hyperfine coupling is known to vary among the compounds of a nuclei due to the varying orbital arrangement but not more than an order of magnitude\cite{abragam}. 
However, for a given material, the hyperfine coupling is known to have a well defined value, which helps the assessment of NMR data in
materials especially when other factors, such as temperature or doping dependence, come into play.
We show that in Weyl semimetals (see Fig. \ref{nmrabra}), the opposite is true: the hyperfine coupling depends strongly on the chemical potential and the temperature. We also calculate the  NMR spin-lattice relaxation time, $T_1$,
 and show that the contribution of Weyl quasiparticles to $T_1$ is negligible. 
 However, the orbital hyperfine coupling itself can be very large and gate controllable, which is highly important for possible application of Weyl semimetals in quantum computing and spintronics.

\section{Nuclear spins in Weyl semimetals}

The Hamilton operator of Weyl semimetals is written as
\begin{equation}
H=v_{\text{F}}(p_x\sigma_x+p_y\sigma_y+p_z\sigma_z),
\label{hamilton}
\end{equation}
where $\sigma$'s are spin-1/2 Pauli matrices, corresponding to the physical spin, $v_{\text{F}}$ is their Fermi velocity, typically\cite{neupane,chiral2} of the order of $10^{5}-10^{6}$~m/s.
Eq. \eqref{hamilton} describes a monopole in momentum space.
Its dispersion relation is also linear in momentum, as is usual for zero mass Weyl fermions in arbitrary dimension (i.e. for graphene as well\cite{rmpguinea}) as
\begin{equation}
\varepsilon_\pm({\bf k}) = \pm v_{\text{F}}\hbar\vert{\bf k}\vert.
\label{weylenergy}
\end{equation}
To simplify notations, we use $k=|\bf{k}|$ for the length of the 3D momentum.
The eigenfunctions are written as
\begin{subequations}
\begin{gather}
\phi_{{\bf k},+}({\bf r})=\frac{1}{\sqrt{V}}\exp(i{\bf kr})\begin{bmatrix} \cos{\left(\frac{\vartheta_{\bf k}}{2}\right)} \\\sin{\left(\frac{\vartheta_{\bf k}}{2}\right)}
\exp(i\varphi_{\bf k}) \end{bmatrix}
\label{weylfunction1}\\
\phi_{{\bf k},-}({\bf r})=\frac{1}{\sqrt{V}}\exp(i{\bf kr})
\begin{bmatrix}\sin{\left(\frac{\vartheta_{\bf k}}{2}\right)} \\-\cos{\left(\frac{\vartheta_{\bf k}}{2}\right)}\exp(i\varphi_{\bf k}) \end{bmatrix},
\label{weylfunction2}
\end{gather}
\label{weylfunction}
\end{subequations}
corresponding to positive and negative eigenenergies, respectively, and spherical coordinates were used such that $\varphi_{\bf k}$ is the azimuthal angle in the ($k_x$,$k_y$) plane and $\vartheta_{\bf k}$ is the polar angle made from the $k_z$ axis, $V$ is
the real space volume of the sample.

We follow the standard route in Refs. \onlinecite{abragam,doranmr1} to obtain the hyperfine interaction. As a first step, the nuclear spin is a represented as a dipole with dipole moment ${\bf m}=\hbar\gamma_{\text{n}}{\bf I}$, whose vector potential is
\begin{gather}
{\bf A} = \frac{\mu_0}{4\pi}\frac{{\bf m}\times {\bf r}}{r^3}=\frac{\mu_0}{4\pi}\hbar\gamma_{\text{n}}\frac{{\bf I}\times {\bf r}}{r^3}=\frac{\mu_0}{4\pi}\nabla\times\left(\frac{{\bf m}}{r}\right).
\label{vectorpot}
\end{gather}
Here $\mu_0$ is the vacuum permeability and $\gamma_{\text{n}}$ is the gyromagnetic ratio of the studied nucleus. This vector potential 
enters into the Hamiltonian through the Peierls substitution as ${\bf p} \rightarrow {\bf p}-e \bf A$, and its magnetic field, $\nabla\times \bf  A$ through the Zeeman term. 

To set the stage for the Weyl case, 
we re-investigate here the case of free electrons, obeying the conventional Schr\"odinger equation, in order to appreciate the changes in the hyperfine interactions afterwards.
For conventional free electrons\cite{abragam}, the hyperfine
interaction is determined from
\begin{gather}
H=\frac{1}{2m}\left({\bf p}-e{\bf A}\right)^2+g\mu_{\text{B}} {\bf S}\nabla\times{\bf A}=\nonumber\\
=\frac{1}{2m}\left({\bf p}-e{\bf A}\right)^2+g\mu_{\text{B}}\frac{\mu_0}{4\pi}{\bf S}\nabla\times\nabla\times\left(\frac{{\bf m}}{r}\right),
\label{hfi1}
\end{gather}
and expanding it to first order in $m$ (here $g\approx2$ is the electron $g$-factor and $\mu_{\text{B}}$ is the Bohr-magneton).
After some standard algebraic manipulation, the conventional form of the hyperfine interaction is recovered as $H_{\text{HFI}}=H_{\text{HFI}}^{\text{orb}}+H_{\text{HFI}}^{\text{spin}}$ with
\begin{gather}
H_{\text{HFI}}^{\text{orb}}=\frac{\mu_0}{4\pi}\hbar\gamma_{\text{n}}g\mu^*{\bf I}\frac{{\bf r}\times {\bf p}}{\hbar r^3},\\
H_{\text{HFI}}^{\text{spin}} =\frac{\mu_0}{4\pi}\hbar\gamma_{\text{n}}g\mu_{\text{B}}{\bf I}\left(\frac{{\bf S}r^2-3{\bf r}\left({\bf Sr}\right)}{r^5}-\frac{8\pi}{3}{\bf S}\delta({\bf r})\right).
\label{hyperfinspin}
\end{gather}
Here, the first term describes the interaction of the nuclear spin with the angular momentum of the surrounding electrons, the second one stems from the spin-dipole interactions and the third one is the Fermi contact term, accounting for the probability
of finding conduction electrons at the position of the nucleus. Here, $\mu^*=m/m^*\mu_{\text{B}}$ is the orbital magnetic moment, which considers the the effective mass,
${\bf S}$ is the conduction electron spin at position $\bf r$. 

In the case of Weyl semimetals, similar considerations yield
\begin{equation}
{ H}=v_{\text{F}}\boldsymbol{\sigma}({\bf p}-e{\bf A})+H_{\text{HFI}}^{\text{spin}}.
\label{hweyl}
\end{equation}
This allows us the deduce the hyperfine interaction in real space form as 
\begin{gather}
H_{\text{HFI}}=\frac{\mu_0}{4\pi}\hbar\gamma_{\text{n}} e v_{\text{F}}{\bf I}\frac{{\bf r}\times \boldsymbol{\sigma}}{r^3}+H_{\text{HFI}}^{\text{spin}},
\label{weylhyperfine}
\end{gather} 
which is our first main result. While the spin-dipole part is identical to that in Eq. \eqref{hyperfinspin},
the orbital part of the hyperfine 
interaction differs significantly from those in normal metals.
In particular, although the latter describes the interaction between nuclear spins and the orbital motion of Weyl fermions, it still contains the Weyl's physical spin 
$\boldsymbol{\sigma}$, thus it also ends up being a spin-spin interaction.

\section{Matrix elements of the hyperfine interaction}

The determination of the nuclear spin relaxation rate involves the matrix elements of the
hyperfine coupling with respect to the eigenfunctions of Weyl fermions in Eqs. \eqref{weylfunction}.
Since a nuclear spin is localized in real space to the nucleus, it induces momentum scattering as well as spin scattering
for the conduction electrons. The required matrix elements read as 
\begin{equation}
\langle{\phi_{{\bf k'},\alpha'}}\vert{H}_{\text{HFI}}\vert{\phi_{{\bf k},\alpha}}\rangle=\int d^3{\bf r} \phi_{{\bf k'},\alpha'}^*({\bf r})
{H}_{\text{HFI}}\phi_{{\bf k},\alpha}({\bf r})
\label{matrix1}
\end{equation}
where  $\alpha$ and $\alpha'$ are $\pm$
and denote the band index.

The eigenfunction in Eqs. \eqref{weylfunction} contain plane waves (i.e. $\exp(i{\bf kr})$) 
for their spatial dependence and a wavevector dependent spinor part, corresponding to the 
nontrivial topology of the Weyl point.
The operations\cite{dorapssb} in Eq. \eqref{matrix1} thus involve a Fourier transformation using the plane waves and matrix-vector multiplications stemming from the spinor part of the
wavefunction.

We  first Fourier transform ${H}_{\text{HFI}}$,
yielding  $\hat{H}_{\text{HFI}}$,  which will depend on the momentum transfer between the incoming ($\bf k$) and outgoing ($\bf k'$) 
electron, ${\bf q}={\bf k}-{\bf k'}$. The action of the spinor part of the wavefunction on the matrix element will be considered in the following section.
The details of the Fourier transform of Eq. \eqref{weylhyperfine} are given in the Appendix.
Using ${\bf S}=\frac{\boldsymbol{\sigma}}{2}$, the Fourier transform of the hyperfine interaction reads after some algebraic manipulation as
\begin{gather}
 \hat{H}_{\text{HFI}}=\frac{\mu_0}{q^2}\gamma_{\text{n}}\hbar{\bf I}\left[ev_{\text{F}}\left(\boldsymbol{\sigma}\times{\bf q}\right)+\frac{g\mu_{\text{B}}}{2}\left({\bf q}\times \left({\bf q}\times \boldsymbol{\sigma}\right)\right)\right].
\label{hfift}
\end{gather}
This allows us to estimate the order of magnitude of the hyperfine coupling in Weyl semimetals: by keeping only the orbital term,
we obtain $\mu_0\gamma_{\text{n}} e\hbar^2 v_{\text{F}}^2/{V_{\text{c}}|\mu|}$, which agrees with the more refined value in Eq. \eqref{ahf}.
It is important to note that for small momentum scattering, the $\hat{H}_{\text{HFI}}$ diverges as $ev_{\text{F}}/q$ for $q\rightarrow 0$ in the orbital part
of the hyperfine coupling. Even when the spinor part of the wavefunction is considered later on, this divergence of the coupling
remains present and will modify the scaling of the relaxation rate in an essential way.
This is in sharp contrast to the case of graphene, where the absolute value of the orbital part of the hyperfine coupling is bounded.
The terms containing $g\mu_{\text{B}}$ remain finite in the same small $q$ limit, since the $1/q^2$ prefactor in Eq. \eqref{hfift} is compensated in the numerator.

\section{The NMR relaxation rate due to Weyl fermions}

In a typical NMR experiment, 
the nuclear Larmor frequency, $\omega_0=B/\gamma_{\text{n}}$ is the smallest energy scale of the problem
due to the heavy mass of the nucleus, $B$ the strength of a small external magnetic field.
Without loss of generality, we also assume that the chemical potential, $\mu$ cuts into the lower energy band,
and $\hbar\omega_0\ll k_{\text{B}}T,\mu$.

The spin relaxation rate measures the changes in the state of the surrounding electrons due to flipping the nuclear spin.
Therefore, $\hat{H}_{\text{HFI}}$ in \eqref{hfift} is rewritten in a more suggestive form as
\begin{equation}
{\hat H}_{\text{HFI}}=I_z h_{\text{HFI}}^z+\frac{1}{2}\left(I_+ h_{\text{HFI}}^- +I_- h_{\text{HFI}}^+\right),
\label{hfirelax}
\end{equation}
where  $I_\pm=I_x\pm iI_y$ and $h_{\text{HFI}}^{z,\pm}$ are $2\times 2$ matrices from Eq. \eqref{hfift}, accounting for the electronic
degrees of freedom.
Using Fermi's golden rule, the lifetime of the nuclear spin
is\cite{abragam,winter,SlichterBook}
\begin{gather}
\frac{1}{T_1}=\frac{\pi}{4\hbar}\sum_{\delta=\pm}\int \frac{d^3 \bf k}{(2\pi)^3}\int \frac{d^3 \bf k'}{(2\pi)^3}
\left|\left\langle k'\left|h_{\text{HFI}}^{-}\right|k\right\rangle\right|^2\times\nonumber\\
\times
\cosh^{-2}\left(\frac{\varepsilon_\delta({\bf k})-\mu}{2k_{\text{B}} T}\right)\delta\left[\varepsilon_\delta({\bf k})-\varepsilon_\delta({\bf k'})+\hbar\omega_0\right],
\label{relaxtfun}
\end{gather}
where $|k\rangle=[\sin{\left(\frac{\vartheta_{\bf k}}{2}\right)},-\cos{\left(\frac{\vartheta_{\bf k}}{2}\right)}\exp(i\varphi_{\bf k})]^T$
is the spinor part of the wavefunction in the lower band.
The very same matrix elements characterizes the upper band as well,
and $h_{\text{HFI}}^{-}$ described a nuclear spin  flip process.

If the matrix element in Eq. \eqref{relaxtfun} is constant for $\bf k\sim k'$, which is the case conventionally, then $1/T_1T\sim \max[k_{\text{B}}T,\mu]^4$.
However,  the matrix element has two unusual features: first
 $\left|\left\langle k'\left|h_{\text{HFI}}^{-}\right|k\right\rangle\right|^2$ scales as $|{\bf k-k}'|^{-2}$ for $\bf k\rightarrow k'$, and its explicit
form is given in the Appendix.
Second, for $k=k'$ and fixed $\bf k$ and $\bf k'$ angle, it diverges as $k^{-2}$ with decreasing $k$ as the Weyl point is approached.
In Eq. \eqref{relaxtfun}, six dimensional integration awaits. 
By changing to spherical coordinates in both $\bf k$ and $\bf k'$, the integral containing the Dirac-delta is performed easily as its argument depends
only on $k$ and $k'$, i.e. on the absolute values.
Due to the smallness of $\omega_0$, it is set to zero everywhere
except for the denominator of the matrix element, 
which contains a $q^4$ term from Eq. \eqref{hfift}. For small momentum scattering, it is would vanish, causing a singularity in the integral, which is cured by retaining a finite $\omega_0$ here.

After performing the $k'$ integral, the term in the denominator takes the form
\begin{gather}
q^2\approx k_0^2+2k^2\left[1-\sin\vartheta\sin\vartheta'\cos(\varphi-\varphi')-
 \cos\vartheta\cos\vartheta'\right],
\label{qnegyzet}
\end{gather}
where $k_0=\frac{\omega_0}{v_{\text{F}}}$ is the Larmor wavenumber, and only the lowest order term in $k_0$ is kept.
The resulting expression is always positive and the divergence at $k\rightarrow 0$ is cut off by the Larmor frequency term.

\begin{figure}[th]
\psfrag{x}[t][][1][0]{$x=k_0/k$}
\psfrag{y}[b][t][1][0]{$F_2(x)$, \color{red} $F_3(x)$, \color{blue} $F_1(x)$}
\includegraphics[width=6.5cm]{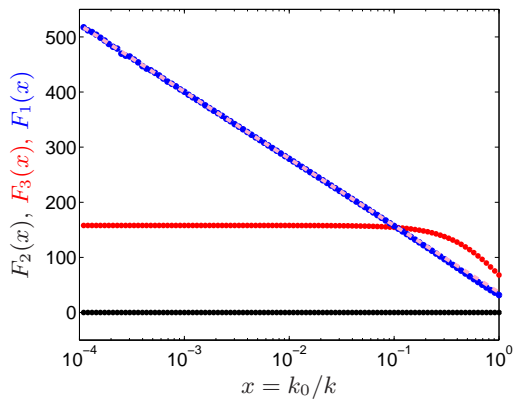}
\caption{(Color online) The numerical evaluation of the $F_{1,2,3}(x)$ (blue, top, black, bottom, red, middle, respectively) functions. The best fitting function for $F_1(x)$ is of the form
$c_1\ln(c_2/x)$ with $c_1\approx 52.7$ and $c_2\approx 2$ (pink dashed line).  Since there is no other scale in the problem, we expect our fitting function to hold down to $x\rightarrow 0$. }
\label{functiona}
\end{figure}

After some algebra, Eq. \eqref{relaxtfun} reduces to 
\begin{gather}
\frac{1}{T_1}=\frac{\pi\mu_0^2\gamma_{\text{n}}^2}{4 v_{\text{F}}(2\pi)^6}\int_{-\infty}^{\infty}dk k^2 \cosh^{-2}\left(\frac{\hbar v_{\text{F}} k-\mu}{2k_{\text{B}} T}\right)\times\nonumber\\
 \times\left( \left(ev_{\text{F}}\right)^2 F_1\left(\frac{k_0}{|k|}\right)+ev_{\text{F}}\frac{g\mu_{\text{B}}}{2}F_2\left(\frac{k_0}{|k|}\right)|k|\right.+\nonumber\\
+\left.\left(\frac{g\mu_{\text{B}}}{2}\right)^2 F_3\left(\frac{k_0}{|k|}\right)k^2 \right)
\label{weylrelax},
\end{gather}
where the dimensionless $F_{1,2,3}(x)$ functions still involve four angular integrals and are given in the Appendix.
The integrals are performed numerically using Monte-Carlo sampling.
The $F_1(x)$ function diverges logarithmically with vanishing $x$, as shown in Fig. \ref{functiona},
It is well fitted by  $F_1(x\rightarrow 0)\approx 52.7 \ln\left(2/{x}\right)$,
while the other two integrals take on a constant value, therefore the $\omega_0\rightarrow 0$ limit can safely be taken.
Upon using scaling with the number of Monte-Carlo steps, $F_2(x)=0$ and $F_3(x\rightarrow 0)\approx 158$ is found, as also visualized in Fig. \ref{functiona}.
By $k$-power counting, the $F_1(x)$ term contains the lowest $T$ power, thus is the most dominant at low temperatures, where only
the low energy dynamics around the Weyl point matters.

Keeping only the dominant term and performing the remaining integrals, we eventually obtain 
\begin{gather}
\frac{\hbar}{T_1k_{\text{B}}T}=\frac{52.7\pi\mu_0^2\gamma_{\text{n}}^2 e^2}{(2\pi)^6 v_{\text{F}}^2}\times\nonumber\\
\times\left\{
\begin{array}{cc}
\left(\dfrac{k_{\text{B}} T}{\hbar}\right)^2 \dfrac{\pi^2}{6}\ln\left(\dfrac{4 k_{\text{B}} T}{\hbar\omega_0}\right),& \textmd{ } \mu\ll k_{\text{B}}T \\
\left(\dfrac{\mu}{\hbar}\right)^2\ln\left(\dfrac{2 \mu}{\hbar\omega_0}\right),& \textmd{ } \mu\gg k_{\text{B}}T.
\end{array}\right.
\label{t1fin}
\end{gather}
These are valid at low temperatures and small chemical potential (i.e. smaller than the bandwidth).

\section{The hyperfine coupling and Overhauser field}

Since our electronic system consist of non-interacting fermions, the conventional Korringa relation \cite{alloul} 
between the relaxation rate and the density of states (DOS) is expected to be recovered, namely
$\left(T_1 T\right)^{-1}=\left(\frac{\pi k_{\text{B}}}{\hbar}\right)A_{\text{hf}}^2\rho(\mu)^2$. 
The DOS for Weyl semimetals is $\rho(E)=V_{\text{c}}E^2/2\pi^2(\hbar v_{\text{F}})^3$ with $V_{\text{c}}$
is the volume of the unit cell. By introducing an effective, energy dependent hyperfine coupling as 
\begin{gather}
A_{\text{hf}}(\mu)=\sqrt{\frac{52.7}{8}} \frac{\mu_0\gamma_{\text{n}} e\hbar^2 v_{\text{F}}^2}{\pi V_{\text{c}}|\mu|},
\label{ahf}
\end{gather}
 the above relation is satisfied. This identification of the hyperfine coupling is further justified by comparing to Eq. \eqref{hfift}, with which it agrees apart
from the numerical prefactor.
 This means that the hyperfine coupling in Weyl semimetals is tunable by doping or gate voltage.
For large velocity and gyromagnetic ratio (17~MHz/T for $^{31}$P) and small unit cell and doping or temperature, it can be of the order of 100~$\mu$eV.
Close to the Weyl point, the hyperfine coupling is sizeable  but the DOS is vanishingly small, while away from the Weyl point, the DOS is enhanced significantly at the expense
of
reducing the hyperfine coupling. This suggests that the nuclear spins are \emph{not} relaxed through Weyl fermions but by some other, non-intrinsic mechanism.
Weyl semimetals often contain  NMR active nuclei (e.g. P, Nb, Ta) with very high natural abundance, and at low energies, 
The coupling in Eq. \eqref{ahf} predicts a strongly enhanced Overhauser field between
the nuclear and electron spins, which is tunable by gate voltages. Such tunability can be useful in controlling coherence in quantum dot devices
 containing Weyl fermions
for quantum information or spintronical devices.
In particular, 
the lifting of the spin blockade in a double quantum dot device 
by Overhauser fields\cite{nazarov} can be manipulated by the gate tunability of the hyperfine fields
of Weyl systems.

Besides the logarithmic term, Eq. \eqref{t1fin} resembles closely to the nuclear spin relaxation time in graphene\cite{doranmr1,dorapssb}, where
the same $T$ and $\mu$ powers  arise from the linearly vanishing DOS in 2D. As opposed to that, the DOS
in Weyl semimetals varies with the square of the energy and its interplay with the diverging hyperfine interaction 
produces a graphene like spin relaxation time with additional log-corrections.
A similar logarithmic Larmor frequency dependence arises in the Hebel-Slichter NMR peak in s-wave superconductors\cite{tinkham} or
in density waves\cite{maniv} due to the divergence of the density of states at the gap edge.
A constant hyperfine coupling, coming from the spin-dipole term (the $C$ function in Eq. \eqref{weylrelax}), produces indeed a
subleading $1/T_1T\sim \max[k_{\text{B}}T,\mu]^4$  scaling.

Finally we comment on the Knight shift, i.e. the shift of the position of the magnetic resonance signal. Neglecting the orbital effect
of the magnetic field on Weyl fermions, as we have done throughout this paper, a Zeeman term, $B_z\sigma_z$ should be added to Eq. \eqref{hamilton}.
The  effect of $B_z$ on $\left\langle k'\left|h_{\text{HFI}}^{-}\right|k\right\rangle$
within first order perturbation theory determines the Knight shift.
However, the magnetic field shifts the Weyl node in the momentum-space by an amount $B_z/v_{\text{F}}$, so that the spin density remains unchanged.
This is analogous to the vanishing  spin susceptibility of Weyl fermions\cite{koshino} within the realm of the low energy theory, Eq. \eqref{hamilton}.

Let us note that the NMR response of a nuclear spin usually resembles closely to the behaviour of a magnetic impurity in a metallic host at high temperatures,
well above the Kondo temperature. This originates from the fact that in both cases, the hyperfine interaction and the Heisenberg exchange term are represented by
a constant coupling. In the present case, however, this mapping ceases to be exact due to the peculiar divergence of the orbital part of the hyperfine interaction.

\section{Conclusions}
We have focused on the hyperfine interaction in Weyl fermions, and the ensuing NMR dynamics. 
While the spin-dipole part of the coupling behaves conventionally as in other metals,
the orbital contribution is found to be tunable by gating or doping the system and 
diverges anomalously at the vicinity of the Weyl point with the inverse energy.
This promises to be relevant for controlling the lifting of the spin blockade in double quantum dot devices\cite{nazarov}.
The spin lattice relaxation time behaves as $1/T_1T\sim E^2\ln(E/\omega_0)$ with $\omega_0$ the nuclear Larmor frequency and $E=\max[\mu,k_{\text{B}}T]$.
This a) differs from naive expectation by an $E^2$ factor due to the anomalous orbital hyperfine coupling in Weyl systems, and
b) is logarithmically enhanced by the Larmor frequency. This resembles to the scaling of the Hebel-Slichter peak in s-wave superconductors.

\begin{acknowledgments}

BD is supported by the Hungarian Scientific Research Fund
Nos. K101244, K105149, K108676.
\end{acknowledgments}

\appendix

\section{The Fourier transform of the hyperfine coupling}

We Fourier transform Eq. \eqref{weylhyperfine}  term by term.
The first term, originating from the interaction between the nuclear spin
and the orbital motion of the electron, involves ${\cal F}\left[ \frac{{\bf r}}{r^3}\right]({\bf q})$, where $\cal F$ denotes the Fourier transform as
\begin{equation}
{\cal F}\left[\frac{{\bf r}}{r^3}\right]({\bf q})=\int d^3{\bf r} \exp(i{\bf qr})\frac{{\bf r}}{r^3}.
\label{four1}
\end{equation}
This is calculated after realizing that the integrand, $\frac{{\bf r}}{r^3}$
is the negative gradient of $\frac{1}{r}$, i.e. the Coulomb potential.
After partial integration, we are left with
\begin{equation}
{\cal F}\left[\frac{{\bf r}}{r^3}\right]({\bf q})=\frac{4\pi i{\bf q}}{q}\int_{0}^{\infty}\sin\left(qr\right) dr
\label{four2},
\end{equation}
where $q=|{\bf q}|$.
Analogously to the Fourier transform of the Coulomb interaction in 3D from the Yukawa potential,
this integral is evaluated as $\lambda\rightarrow0$ limit of
\begin{equation}
{\cal F}\left[\frac{{\bf r}}{r^3}\right]({\bf q})=\frac{4\pi i{\bf q}}{q}\lim_{\lambda\rightarrow 0}\int_{0}^{\infty}e^{-\lambda r}\sin\left(qr\right)dr,
\label{four3}
\end{equation}
which yields
\begin{equation}
{\cal F}\left[\frac{{\bf r}}{r^3}\right]({\bf q})=\frac{2\pi {\bf q}}{iq}\lim_{\lambda\rightarrow 0}\frac{-2iq}{q^2+\lambda^2}=-\frac{4\pi{\bf q}}{q^2}.
\label{four4}
\end{equation}
Similarly to how the Fourier transform of the Coulomb interaction behaves in various dimensions\cite{giamarchi},
the graphene case\cite{dorapssb} in 2D contains only a single $q$ term in the denominator of Eq. \eqref{four4}.

The Fourier transform of the Zeeman term proceeds along similar steps.
The spin dipole term can be rewritten in terms of directional derivatives as \cite{nufft}
\begin{gather}
{\cal F}\left[\frac{{\bf IS}r^2-3({\bf Ir})({\bf Sr})}{r^5}\right]({\bf q})=\nonumber\\
={\cal F}\left[-\frac{4\pi}{3}({\bf IS})\delta({\bf r})-\partial_{{\bf IS}}\left(\frac{1}{3r}\right)\right]({\bf q})=\nonumber\\
=-\frac{4\pi}{3}({\bf IS})+\frac{4\pi ({\bf Sq})({\bf Iq})}{q^2},
\label{fourzee1}
\end{gather}
where $\partial_{{\bf IS}}=({\bf I}\nabla)({\bf S}\nabla)$ is the directional derivative.
The Fourier transform of the last term containing the Dirac-delta function gives trivially one.

\section{The matrix element of nuclear spin flip}

The matrix element, appearing in Eq. \eqref{relaxtfun}
is obtained by selecting only those terms from Eq. \eqref{hfift}, which contain the 
$x$ and $y$ components of the nuclear spin, giving  $h_{\text{HFI},x}$ and $h_{\text{HFI},y}$.
These define  $h_{\text{HFI}}^-=h_{\text{HFI},x}-ih_{\text{HFI},y}$, which eventually yields
\begin{gather}
\langle{\phi_{{\bf k'},-}}\vert h_{\text{HFI}}^-\vert{\phi_{{\bf k},-}}\rangle=\frac{\mu_0}{q^2}\gamma_{\text{n}}\hbar\left(ev_{\text{F}} F_e+\frac{g\mu_{\text{B}}}{2}F_g\right),
\end{gather}
where in spherical coordinates, we have
\begin{gather}
k_x=k\sin(\vartheta_{\bf k})\cos(\varphi_{\bf k}),\\
k_y=k\sin(\vartheta_{\bf k})\sin(\varphi_{\bf k}),\\
k_z=k\cos(\vartheta_{\bf k}),
\end{gather}
and similarly for $\bf k'$ with the restriction $k'=k+k_0$ due to the Dirac delta in Eq. \eqref{relaxtfun}, and $\bf q=k'-k$.
Additionally,
\begin{gather}
F_e=q_z(s_y+is_x)-s_z(q_y+iq_x),\\
F_g=(q_y+iq_x)(s_yq_x-q_ys_x-iq_zs_z)-q_z^2(s_x-is_y).
\end{gather}
We have also defined 
\begin{gather}
s_i=\left\langle k'|\sigma_i|k\right\rangle,
\end{gather}
$i=x,y,z$, $|k\rangle=[\sin{\left(\frac{\vartheta_{\bf k}}{2}\right)},-\cos{\left(\frac{\vartheta_{\bf k}}{2}\right)}\exp(i\varphi_{\bf k})]^T$ and similarly for $|k'\rangle$.
In particular, 
\begin{gather}
s_x=-\sin\left(\frac{\vartheta_{\bf k}}{2}\right)\cos\left(\frac{\vartheta_{\bf k'}}{2}\right) \exp(-i\varphi_{\bf k'})-\nonumber\\
-\sin\left(\frac{\vartheta_{\bf k'}}{2}\right)\cos\left(\frac{\vartheta_{\bf k}}{2}\right) \exp(i\varphi_{\bf k}),\\
s_y=-i\sin\left(\frac{\vartheta_{\bf k}}{2}\right)\cos\left(\frac{\vartheta_{\bf k'}}{2}\right) \exp(-i\varphi_{\bf k'})+\nonumber\\
+i\sin\left(\frac{\vartheta_{\bf k'}}{2}\right)\cos\left(\frac{\vartheta_{\bf k}}{2}\right) \exp(i\varphi_{\bf k}),\\
s_z=\sin\left(\frac{\vartheta_{\bf k}}{2}\right)\sin\left(\frac{\vartheta_{\bf k'}}{2}\right)-\nonumber\\
-\cos\left(\frac{\vartheta_{\bf k}}{2}\right)\cos\left(\frac{\vartheta_{\bf k'}}{2}\right) \exp(i(\varphi_{\bf k}-\varphi_{\bf k'})).\\
\end{gather}

\section{The $F_1(x)$, $F_2(x)$ and $F_3(x)$ functions}

First, we define the auxiliary functions
\begin{gather}
a=\frac{k^2|F_e|^2}{q^4},\\
b=\frac{2k\textmd{Re}[F_eF_g^*]}{q^4},\\
c=\frac{|F_g|^2}{q^4}
\end{gather}
with $q^2$ from Eq. \eqref{qnegyzet}.
By multiplying them 
 with $\sin(\vartheta_{\bf k})\sin(\vartheta_{\bf k'})$, stemming from the Jacobian,
and integrating them with respect to the four angular variables $\vartheta_{\bf k}$, $\vartheta_{\bf k'}$ from 0 to $\pi$ and
$\varphi_{\bf k}$, $\varphi_{\bf k'}$
 from 0 to $2\pi$, 
we get the desired $F_1(x)$, $F_2(x)$ and $F_3(x)$ functions.
The resulting dimensionless functions depend only on the ratio $k_0/k$, which is denoted by $x$, and not separately on $k_0$ and $k$.

\bibliographystyle{apsrev}

\bibliography{graph_nmr}

\end{document}